\documentclass[12pt]{iopart}
\usepackage{iopams}

\newcommand{\be}{\begin{equation}}
\newcommand{\ee}{\end{equation}}
\newcommand{\ben}{\begin{eqnarray}}
\newcommand{\een}{\end{eqnarray}}
\newcommand{\ra}{\rangle}
\newcommand{\la}{\langle}
\newcommand{\nn}{\nonumber}

\usepackage{iopams,epsf}  
\begin{document}
\title[Influence of the spin quantum number  on zero-temperature phase 
transition]
{Influence of the spin quantum number $s$ on the zero-temperature phase 
transition in the square lattice $J$-$J'$  model}
\author
{
R.Darradi$^a$, J.Richter$^a$, and D. J. J. Farnell$^b$
}
\address
{
$^a$Institut f\"ur Theoretische Physik, Otto-von-Guericke Universit\"at
Magdeburg, \\
P.O.B. 4120, 39016 Magdeburg, Germany \\   
$^b$Unit Of Ophthalmology, Department of Medicine,
University Clinical Departments, Daulby Street,
University of Liverpool, Liverpool L69 3GA, United Kingdom\\
}


\begin{abstract}
We investigate the phase diagram of the Heisenberg antiferromagnet 
on the square lattice with two different nearest-neighbor bonds 
$J$ and $J'$ ($J$-$J'$ model) at zero temperature.
The model exhibits a quantum phase transition at a critical value $J'_c > J$
between  a semi-classically ordered N\'eel and a magnetically disordered 
quantum paramagnetic phase of valence-bond type, which  
is driven by local singlet formation on $J'$
bonds. We study the influence of spin quantum number $s$ on this 
phase transition by means of a variational mean-field 
approach, the coupled cluster method, and the Lanczos 
exact-diagonalization technique. 
We present evidence that  the critical  value $J'_c$ increases with
growing $s$ according to $J'_c \propto s(s+1)$.
\end{abstract}

\maketitle


\section{Introduction}
The study of quantum antiferromagnets in low-dimensional 
systems has attracted much attention 
in recent years, both theoretically and experimentally. 
In particular, quantum phase transitions  are in the 
focus of
interest, see e.g. Refs. 
\cite{sachdev,qpt_ri01,sachdev04}. For these zero-temperature transitions  thermal fluctuations are 
irrelevant and the transition between different quantum phases  (e.g.
between
magnetically ordered  
and disordered phases) is driven purely by quantum fluctuations.
For the quantum spin Heisenberg antiferromagnet (HAFM) 
on two-dimensional lattices the interplay of
interactions and fluctuations is well balanced
and the existence of semi-classical magnetic long-range order depends on the
degree of competition between bonds \cite{lhuillier03,wir04}. 
Competition between bonds in spin systems may appear as  frustration,
which is present in classical as well as in quantum spin systems. 
In quantum systems
also a direct competition between bonds exists which may lead to 
local singlet formation on certain antiferromagnetic bonds 
(or plaquettes of four spins) if
these bonds are increased in strength. 
By tuning the degree of competition zero-temperature 
 order-disorder phase transitions can be realized.  
The
existence of magnetically  disordered
quantum paramagnetic ground states in (quasi-)two-dimensional 
Heisenberg systems has been 
recently demonstrated,  e.g. for
$\mathrm{SrCu}_2(\mathrm{BO}_3)_2$
\cite{taniguchi,Troyer} and
$\mathrm{CaV}_4\mathrm{O}_9$ \cite{Kageyama,Koga}. 

A canonical model to study the competition in a frustrated 
quantum spin HAFM is the  $J_1$-$J_2$ model on the square lattice, 
where the frustrating $J_2$ bonds 
plus quantum fluctuations lead to a second-order transition from 
N\'eel ordering to a quantum paramagnetic phase, 
see e.g. \cite{doucot88,Richter93,Oitmaa96,bishop98,Capriotti02}.
A widely studied model describing competition without frustration and showing
the 'melting' of semi-classical N\'eel 
order by local singlet formation is the HAFM on the square lattice 
with two 
non-equivalent nearest-neighbor bonds 
$J$ and  $J'$ 
($J$-$J'$ model) \cite{sing88,ivanov96,krueger00,ivanov96,
krueger01,tomczak01,matsumoto01,rachid04,yoshioka04}.
In these 
papers on the $J$-$J'$ model
the extreme quantum case $s=1/2$ is considered and the competition can be
tuned by variation of the exchange bond $J'$.
One finds a second-order transition from
the quasi-classically N\'eel ordered phase to a dimerized singlet phase 
at $J'_c \approx 2.5 \ldots 2.9 J$. It is argued 
in Refs. \cite{tomczak01,matsumoto01}
that the quantum phase transition  
is of the same universality class
as the thermal phase transition of 
three-dimensional classical Heisenberg model.

The strength of quantum fluctuations within this model can be varied either
by anisotropy or by spin quantum number. Indeed its was found in
\cite{tomczak01} for the $J$-$J'$ model
that the critical $J'_c$ for the {\it XY} model is significantly 
larger than
for the spin rotationally invariant Heisenberg model. The influence of an
Ising exchange anisotropy $\Delta_I$ leads also to an increase of $J'_c$
which is in good approximation proportional to $\Delta_I$ \cite{rachid04}.
The role of the spin quantum number $s$ was not systematically studied. Some
results for spin models with $s=1$ can be found in e.g.,
\cite{matsumoto01,matsumoto03}

In the present paper we study the ground state phase transition 
between a N\'eel ordered phase and a dimerized singlet phase
of the $J$-$J'$ model with spin quantum number $s=1/2, 1, 3/2, 2$  
using a variational mean-field like 
approach (MFA), the coupled cluster method (CCM) and 
exact diagonalization (ED) of finite systems. 
\section{Model}
\begin{figure}[ht]
\begin{center}\epsfxsize=20pc
\epsfbox{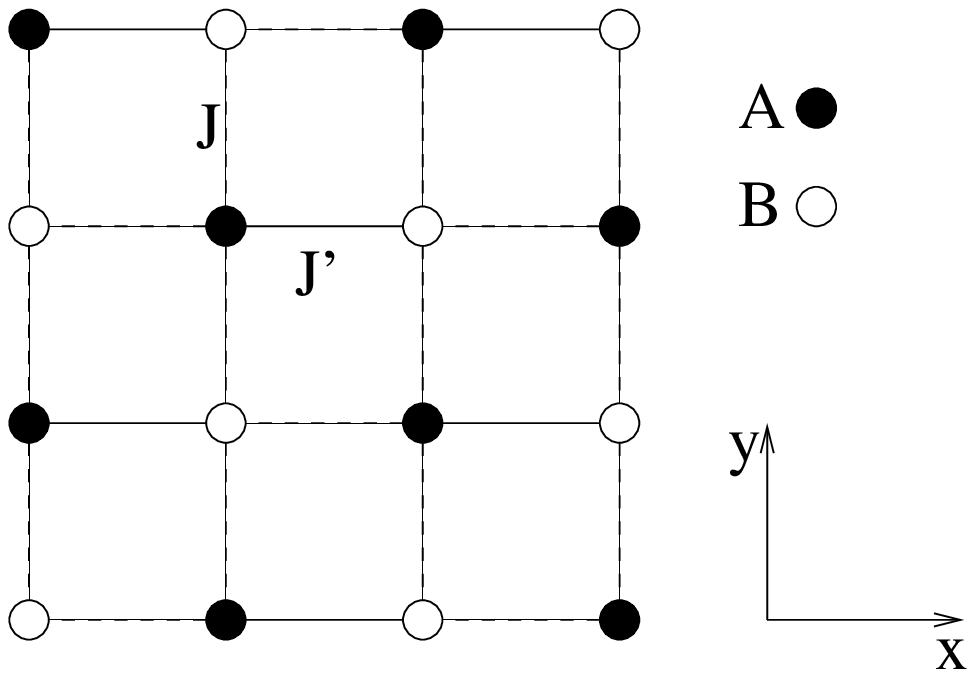}
\end{center}
\caption{\label{fig1} 
Illustration of arrangement of bonds in the 
$J$-$J'$ model on the  square lattice
(Eq.(\ref{eq1})): 
$J$ -- dashed lines;   $J'$ -- solid lines; A and B
characterize the two sublattices of the classical N\'eel ground state.
}
\end{figure}

We consider the $J$-$J'$ model on a square lattice, i.e. a HAFM
with two kinds of antiferromagnetic nearest-neighbor bonds $J$ and $J'$ (see 
Fig.\ref{fig1}) described by the Hamiltonian
\begin{equation}
\label{eq1}
H = J\sum_{<ij>_1}{\bf s}_i \cdot {\bf s}_j
+ J'\sum_{<ij>_2}{\bf s}_i \cdot {\bf s}_j,
\end{equation}
where the sums over $<ij>_1$ and $<ij>_2$ represent sums
over the nearest-neighbor bonds, shown in Fig.\ref{fig1} as dashed and solid lines,
respectively. We consider 
spin operators  
${\bf s}_i^2 = s(s+1)$ of spin quantum number $s = 1/2, 1, 3/2, 2$.\\
Each square-lattice plaquette consists of three $J$ bonds and one $J'$
bond. 
In what follows we set
$J=1$ and consider $J'\ge 1$ as the parameter of the model.
In the classical limit the ground state is the symmetry breaking 
N\'eel state. However, quantum
fluctuations may lead to a rotationally invariant dimerized valence-bond
state for large enough $J'$.
\section{Variational mean-field like approach}\label{mfa}
In this section we use the MFA  
to calculate the ground-state magnetic order parameter of the $J$-$J'$ model
(\ref{eq1}).
For the spin half HAFM this approach has been successfully applied to  
bilayer systems \cite{gros95}, to the isotropic \cite{qpt_ri01,krueger00} 
and
anisotropic \cite{rachid04} $J$-$J'$ model on the square lattice and on the $1/5$ depleted
square lattice for $\mathrm{CaV}_4\mathrm{O}_9$ \cite{wir04}, but also on
the strongly frustrated HAFM on the star lattice \cite{star04}.
In this paper we extent the basic ideas of this approach to   
higher spin quantum numbers $s= 1, 3/2, 2$.

We start with the two uncorrelated mean-field like states, namely  
the N\'eel state 
$|\phi_{MF_1}\rangle  = |\hspace{-0.01cm} +s \rangle
|\hspace{-0.01cm} -s \rangle |\hspace{-0.01cm}+s \rangle|\hspace{-0.01cm}
-s \rangle 
\ldots \;$ and the dimerized  rotationally invariant  
singlet product state (valence-bond state) 
$|\phi_{MF_2}\rangle  = 
\prod_{<ij>_2}  |\{i,j\}_s \rangle$, where the 
product runs over all $J'$ bonds, cf. (\ref{eq1}).
$ |\{(i,j)\}_s \rangle$ in $|\phi_{MF_2}\rangle$ is a singlet state of two
spins $s$, i.e. we have 
$ |\{i,j\}_{s=1/2} \rangle =  \frac{1}{\sqrt{2}}
[ |+\frac{1}{2}\rangle |-\frac{1}{2}\rangle 
- |-\frac{1}{2}\rangle |+\frac{1}{2}\rangle, \;$  
$ |\{i,j\}_{s=1} \rangle =
 \frac{1}{\sqrt{3}}[ |+1\rangle |-1\rangle -|0\rangle 
|0 \rangle + |-1\rangle |+1\rangle ], \;$
$ |\{i,j\}_{s=3/2} \rangle =\frac{1}{2}[ |+\frac{3}{2}\rangle |-\frac{3}{2} \rangle
- |+\frac{1}{2}\rangle |-\frac{1}{2}\rangle + 
|-\frac{1}{2}\rangle |+\frac{1}{2}\rangle
-|-\frac{3}{2}\rangle |+\frac{3}{2} \rangle ], \;$
$ |\{i,j\}_{s=2} \rangle=
\frac{1}{\sqrt{5}}[|+2\rangle |-2 \rangle - |+1\rangle |-1 \rangle + 
|0\rangle |0 \rangle
-|-1\rangle |+1 \rangle + |-2\rangle |+2 \rangle].  \;$

In order to describe the transition between 
both states we consider for the different spin quantum numbers respective 
uncorrelated product trial states of the form
\begin{eqnarray}
\hspace{-1.cm}|\Psi^{s=1/2}_{var}\rangle &=& \prod_{<ij>_2} \frac{1}{\sqrt{1+a^2}}\Bigg [ |+\frac{1}{2}\rangle |-\frac{1}{2}\rangle 
- a|-\frac{1}{2}\rangle |+\frac{1}{2}\rangle\Bigg ]\label{eq2}\\
\hspace{-1.cm}|\Psi^{s=1}_{var}\rangle &=& \prod_{<ij>_2} \frac{1}{\sqrt{1 + b_1^2+b_2^2}}
\Bigg [ |+1\rangle |-1\rangle - b_1|0\rangle |0 \rangle + b_2|-1\rangle |+1\rangle 
\Bigg ]\label{eq2a}\\
\hspace{-1.cm}|\Psi^{s=3/2}_{var}\rangle &=&
 \prod_{<ij>_2} \frac{1}{\sqrt{1+c_1^2+c_2^2+c_3^2}}
\Bigg [ |+\frac{3}{2}\rangle |-\frac{3}{2} \rangle -
c_1|+\frac{1}{2}\rangle |-\frac{1}{2}\rangle  \nonumber\\
&& \qquad  + c_2|-\frac{1}{2}\rangle |+\frac{1}{2}\rangle
-  c_3|-\frac{3}{2}\rangle |+\frac{3}{2} \rangle \Bigg ]\label{eq2b}\\
\hspace{-1.cm}|\Psi^{s=2}_{var}\rangle &=& \prod_{<ij>_2} 
\frac{1}{\sqrt{1+d_1^2+d_2^2+d_3^2+d_4^2}}
\Bigg [|+2\rangle |-2 \rangle - d_1|+1\rangle |-1 \rangle  
  \nonumber\\ 
&& \qquad  + d_2|0\rangle |0 \rangle - d_3|-1\rangle |+1 \rangle 
+ d_4|-2\rangle |+2 \rangle\Bigg ] \label{eq2c},
\end{eqnarray}
where in the two-spin states $|n\rangle |m\rangle$ the first bra vector belongs to site $i$ and the 
second to site $j$ of a $J'$ bond. 
The trial wave functions depend on the variational parameters $a$;\, 
$b_1$, $b_2$;\, $c_1$, $c_2$ $c_3$;\, 
$d_1$, $d_2$, $d_3$, $ d_4$ and interpolate between
 the  valence-bond state
$|\phi_{MF_2}\rangle$  realized for $a = 1$;\, $b_1= b_2 = 1$;\, 
$c_1 = c_2 = c_3 = 1$;\, $d_1 = d_2 = d_3 = d_4 = 1$ and the 
N\'eel state  
$|\phi_{MF_1}\rangle$ for  $a = 0$; \, $b_1 = b_2 = 0$; \, $c_1 = c_2 = c_3 = 0$;\, 
$d_1 =  d_2 =  d_3 =  d_4 = 0$, respectively.
The ground-state energy per site $e^s_{var} =
\langle\Psi^s_{var}|H|\Psi^s_{var}\rangle/N$ is
calculated as
\begin{eqnarray}
\label{eq3}
\hspace{-2.1cm} e^{s=1/2}_{var}(a) &=& -\frac{J'}{2}\;\frac{a+\frac{1}{4}
(1+a^2)}{1+a^2} - \frac{3}{2}\;\frac{(1-a^2)^2}{4(1+a^2)^2} \\
\hspace{-2.1cm}e^{s=1}_{var}(b_1, b_2) &=& -\frac{J'}{2}\;
\frac{2b_1+2b_1b_2+1+b_2^2}{1+b_1^2+b_2^2} - 
\frac{3}{2}\; \frac{(1 - b_2^2)^2}{(1+b_1^2+b_2^2)^2} \\ 
\hspace{-2.1cm}e^{s=3/2}_{var}(c_1, c_2, c_3) &=& -\frac{J'}{2}\;
\frac{3c_1+4c_1c_2+3c_2c_3+\frac{1}{4}
(9+c_1^2+c_2^2+9c_3^2)}{1+c_1^2+c_2^2+c_3^2}\nn  \\
&&- \frac{3}{2}\;\frac{(3+c_1^2-c_2^2-3c_3^2)^2}{4(1+c_1^2+c_2^2+c_3^2)^2} \\
\hspace{-2.1cm}e^{s=2}_{var}(d_1, d_2, d_3, d_4) &=& -\frac{J'}{2}\;
\frac{4d_1+6d_1d_2+6d_2d_3+4d_3d_4+d_1^2
+4+d_3^2+4d_4^2}{1+d_1^2+d_2^2+d_3^2+d_4^2}  \nn\\
&& -\frac{3}{2}\;\frac{(2+d_1^2-d_3^2-2d_4^2)^2}
{(1+d_1^2+d_2^2+d_3^2+d_4^2)^2}.
\end{eqnarray}
The  relevant order parameter describing the N\'eel order is the sublattice 
magnetization $M = \langle \Psi^s_{var}| s^z_{i \in A}|\Psi^s_{var}\rangle$. 
Using Eqs. (\ref{eq2}), (\ref{eq2a}), (\ref{eq2b}), (\ref{eq2c}) we obtain
\begin{eqnarray}
\label{eq5}
M_{s=1/2}(a)  &=& \frac{1-a^2}{2+2a^2} \\
M_{s=1}(b_1, b_2)  &=& \frac{1-b_2^2}{1+b_1^2+b_2^2} \\
M_{s=3/2}(c_1, c_2, c_3)  &=&
\frac{3+c_1^2-c_2^2-3c_3^2}{2(1+c_1^2+c_2^2+c_3^2)}\\
M_{s=2}(d_1, d_2, d_3, d_4) &=&  
\frac{2+d_1^2-d_3^2-2d_4^2}{1+d_1^2+d_2^2+d_3^2+d_4^2} .
\end{eqnarray}

\begin{figure}[ht]
\begin{center}\epsfxsize=30pc
\epsfbox{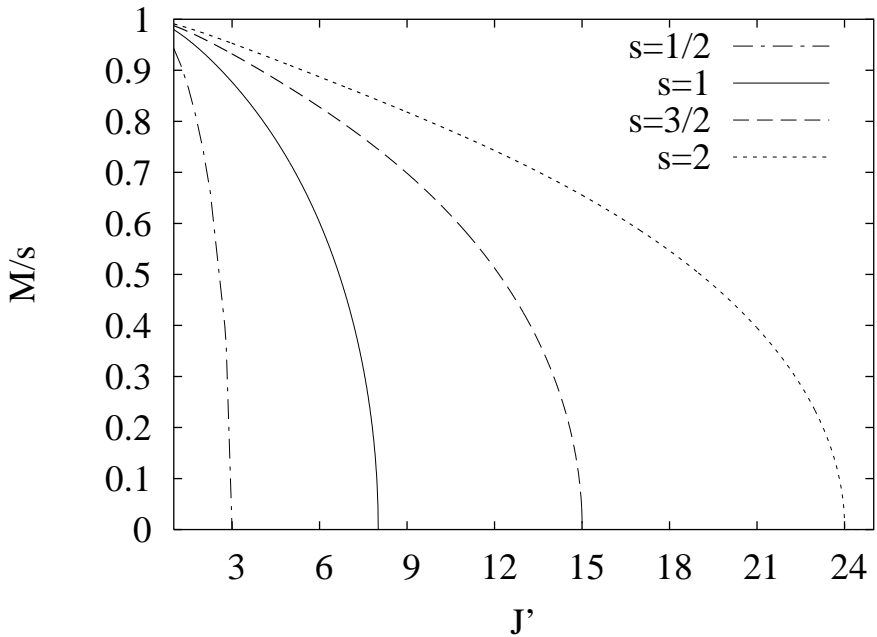}
\end{center}\caption{\label{fig2} Sublattice magnetization $M/s$ versus $J'$    
calculated by the variational mean-field like approach  (MFA), see text.}
\end{figure}
%

We minimize E = $\langle\Psi_{var}|H|\Psi_{var}\rangle$ with respect to 
the variational 
parameters. As a result we obtain an analytic expression for $a$ in the
case of $s=1/2$,
but a set of 2, 3, 4 coupled 
nonlinear equations for $s=1, 3/2, 2$ to determine 
the variational parameters.
As reported in \cite{krueger00,qpt_ri01} 
the sublattice magnetization for $s=1/2$ is 
$ M_{s=1/2}(J') = 
\frac{1}{2}\sqrt{1-(J'/3 )^2}$ for $J'\leq
3$ but zero for $J' > 3$. Furthermore, one can express the ground-sate
energy 
as a Landau-type function of $M$, $e^{s=1/2}_{var}=
-\frac{3}{8}J' +  \frac{1}{2}(J' - 3)M^2  +\frac{1}{2} J'
M^4$, indicating the molecular field-like nature of the
approach.
For $s=1,\, 3/2,\, 2$ we have to solve the corresponding set of nonlinear equations 
numerically.
We show $M(J')$ in  
Fig.\ref{fig2}. 
$M(J')$ vanishes 
at a critical point $J'_c=3$ $(s=1/2)$, $J'_c=8$ $(s=1)$, $J'_c=15$
$(s=3/2)$, 
$J'_c=24$ $(s=2)$, respectively. 
The corresponding  critical index is the mean-field index 
$1/2$.

The sequence of critical points for $s=1/2,\ldots,2$ are precisely described
by  $J_c(s)=\frac{4}{3}s(s+1)(z-1)$, where $z=4$ 
is the coordination number of the square lattice.
Although we do not have results for $s>2$, we argue that due to the systematic
character of the MFA approach  it seems to be likely  that this expression
is valid also for $s> 2$. 
\section{Coupled cluster method (CCM)}\label{ccm}
We now briefly describe the general CCM formalism, for further details the interested reader is 
referred to Refs.
\cite{zeng98,bishop00,farnell01,farnell02}. 
In order to calculate the many-body ground state, we start with 
a normalized reference or model state $|\Phi\rangle$.  
We chose the N\'eel state as the reference state $|\Phi\rangle$ 
in order to treat the $J$-$J'$ model using the CCM.
It is convenient to perform a rotation of the local axis of
the up spins such that all spins in the reference state align in the same
direction, namely along the negative $z$-axis, such that we have 
$\; \; |\Phi\rangle  = |-s \rangle|-s \rangle 
|-s \rangle| -s \rangle
\ldots \; \;$. 
We define a set of multi-spin creation operators $C_I^+=s_r^+ \; , \;
s_r^+ s_l^+ \; , \; s_r^+ s_l^+ s_m^+\; , \; \ldots \;$ . 
With this definition of the $C_I^+$ we have $\langle \Phi| C_I^+ = 0 =
C_I|\Phi\rangle$, where $C_I$ is the Hermitian adjoint of $C_I^+$. 
The CCM  ket and bra ground states are then given by
\be\label{ket} 
  |\Psi\ra =e^S|\Phi\ra, \quad S=\sum_{I\neq 0}{\cal S}_IC_I^+ ,\ee
\be\label{bra} 
  \la\tilde\Psi|=\la\Phi|\tilde Se^{-S}, \quad \tilde S=
1+\sum_{I\neq 0}\tilde {\cal S}_IC_I .\ee
The correlation operators $S$ and $\tilde S$ contain 
the  correlation coefficients
${\cal S}_I$ and $\tilde {\cal S}_I$ which have to be determined.
Using the Schr\"odinger equation, $H|\Psi\ra=E|\Psi\ra$, we can now write 
the ground-state energy as $ 
    E=\la\Phi|e^{-S}He^S|\Phi\ra$ . 
The sublattice magnetization  
is given by $
 M = -\la\tilde\Psi|s_i^z|\Psi\ra$ .

\begin{figure}[ht]
\begin{center}\epsfxsize=30pc
\epsfbox{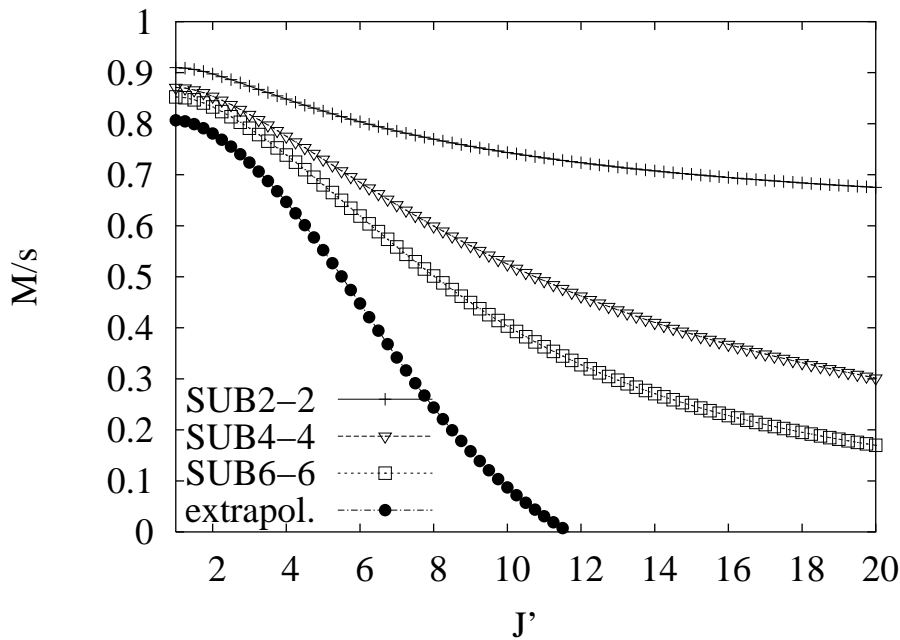}
\end{center}
\caption{\label{fig3} Sublattice magnetization $M/s$ versus $J'$ for 
spin quantum number  $s=1$  
using coupled cluster method  (CCM), see text.}
\end{figure}

In order to determine the  correlation coefficients 
${\cal S}_I$ and $\tilde {\cal S}_I$, we 
to require that the expectation value $\bar H=\la\tilde\Psi|H|\Psi\ra$
is a minimum with respect to 
${\cal S}_I$ and $\tilde {\cal S}_I$.
In case we would be able to take into account  all possible multispin 
configurations in
the correlation operators $S$ and $\tilde S$ the CCM formalism is exact. 
However, for the considered quantum spin model
we have to use approximation schemes to truncate the expansion
of $S$ and $\tilde S$ in the Eqs.~(\ref{ket}) and (\ref{bra}). As in
Refs. \cite{zeng98,bishop00,farnell02}, we use the 
SUB$n\hspace{-0.1cm}-\hspace{-0.1cm}n$ approximation scheme, 
where we include only $n$ spin flips  
in all configurations (or lattice animals in the language of graph theory)
which span a range of no more than $n$ adjacent 
lattice sites. Note that this approximation for $s=1/2$ 
is equivalent to the LSUB$n$ approximation \cite{zeng98,bishop00,farnell02}.
Since the approximation becomes exact in the limit $n \to \infty$ 
is useful to extrapolate the 'raw' CCM-SUB$n\hspace{-0.1cm}-\hspace{-0.1cm}n$ 
results to the limit $n \to \infty$.
Although an exact scaling theory for SUB results is not known, 
there is empirical experience \cite{krueger00,zeng98,bishop00,farnell02}
how the order parameter for antiferromagnetic 
spin models scales with $n$. 
In accordance with those findings we use 
$ M(n)=M(\infty)+a_1(1/n)+a_2(1/n)^2$ to extrapolate to $n \to \infty$.
We note that we take a value of  $M(\infty)$ tending to zero 
to indicate the critical point $J'_c$ (see Fig.\ref{fig3}).
The values for $J'_c$  obtained by extrapolation of the 
SUB$n\hspace{-0.1cm}-\hspace{-0.1cm}n$ results
for $M$ are,
however, found to be slightly too large \cite{krueger00,rachid04}. 
Therefore, it is more favorable to consider the inflection
points of the $M(J')$ curve for 
the SUB$n\hspace{-0.1cm}-\hspace{-0.1cm}n$ approximation, 
assuming that the
true $M(J')$ curve will have a negative curvature up to the critical point. 
For increasing $n$ the inflection point $J'_{inf}$ 
approaches the critical point $J'_c$.
The 
inflection points for the SUB$n\hspace{-0.1cm}-\hspace{-0.1cm}n$ approximation
again we can extrapolate to the limit $n\to\infty$ 
using $ J'_{inf}(n)=J'_{inf}(\infty)+b_1(1/n)+b_2(1/n)^2$
and interpret $J'_{inf}(\infty)$ as the critical value $J'_c$.
%
\begin{figure}[ht]
\begin{center}\epsfxsize=30pc
\epsfbox{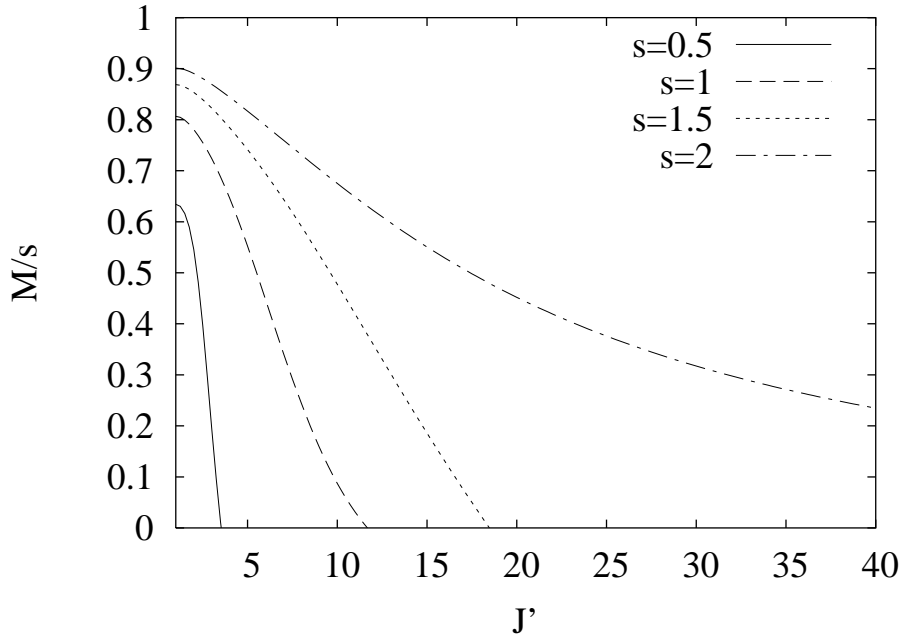}
\end{center}
\caption{\label{fig4} Extrapolated  
sublattice magnetization $M/s$ versus $J'$ for 
various spin quantum numbers  $s$  
using coupled cluster method  (CCM), see text.}
\end{figure}

In principle it is possible to apply the CCM for arbitrary spin
quantum number $s$. However, within the used 
SUB$n\hspace{-0.1cm}-\hspace{-0.1cm}n$ approximation scheme for higher $s$
additional problems appear, namely (i) the number of fundamental
configurations (lattice animals) increases with $s$, which makes the
calculations on a certain level of approximation $n$ 
more ambitious and (ii) the total
number  of basis states  grows drastically with $s$ according to $s^N$
and 
as a consequence the 
SUB$n\hspace{-0.1cm}-\hspace{-0.1cm}n$ approximation may become less
reliable.
While the latter point is irrelevant for systems where the quantum
ground state is close to the reference state (i.e. in our model in case of
well-pronounced N\'eel order) it becomes relevant if the quantum ground state
is far from the reference state (i.e., in our model when N\'eel order breaks
down).
Hence the results for higher spin quantum numbers must be taken with extra 
care.

We have calculated CCM results within the 
SUB$n\hspace{-0.1cm}-\hspace{-0.1cm}n$ approximation  for $n=2,4,6$
for $s=1/2,1,3/2,2$. For spin $1/2$ also results for $n=8$ are available
(see Refs. \cite{krueger00,rachid04}).
First we report the values for $J'_c$ for spin $s=1/2$ (see also 
Refs. \cite{krueger00,rachid04}). 
The extrapolation of the SUB$n\hspace{-0.1cm}-\hspace{-0.1cm}n$ 
data for $M$ with
$n=2,4,6$ as described above leads to $J'_c \approx 3.5$.
However,
as discussed above the extrapolation of the order parameter tends to
overestimate $J'_c$ (note that $J'_c$ obtained this way is even larger
than the value found within MFA) and the extrapolation of the inflection point is
favorable. We found as inflection points of the $M_s(J')$
curves  
$J'_{inf}(n)= 3.60 $  (SUB$2\hspace{-0.1cm}-\hspace{-0.1cm}2$), $3.33 $ 
(SUB$4\hspace{-0.1cm}-\hspace{-0.1cm}4$),   $3.13$  
(SUB$6\hspace{-0.1cm}-\hspace{-0.1cm}6$)
leading to an extrapolated value  of $J'_c= J'_{inf}(\infty)=
 2.56$.
We mention that the consideration of  SUB$8\hspace{-0.1cm}-\hspace{-0.1cm}8$
data leads to a slight modification of 
$J'_c$ to $J'_c= J'_{inf}(\infty) = 2.54$, only.

We now consider the case $s=1$, where the results for the
order parameter $M$ are giving in Fig. \ref{fig3}.
Clearly we see the weakening of the magnetic order by increasing $J'$.
The extrapolation of the SUB$n\hspace{-0.1cm}-\hspace{-0.1cm}n$
data for $M$ with
$n=2,4,6$ leads to $J'_c \approx 11.7$, i.e. we get the same
tendency as for the variational MFA that $J'_c$ increases with $s$. Again
the extrapolation of the order parameter leads to an
overestimation of $J'_c$. 
This overestimation is connected with the change
in the sign of curvature of $M(J')$ seen in Fig. \ref{fig3}. 
 The favorable 
extrapolation of the inflection points 
leads to
 $J'_c= J'_{inf}(\infty) \approx
 6.4$, where the inflection points for the different levels of 
SUB$n\hspace{-0.1cm}-\hspace{-0.1cm}n$ approximations are  
$J'_{inf}(n=2)= 3.93 $, $J'_{inf}(n=4)= 6.04$,  
$J'_{inf}(n=6)= 6.36$.

\begin{table}
\caption{CCM results for the ground state of the 
Heisenberg antiferromagnet  on the square lattice 
with spin quantum number $s=3/2$ and $s=2$
using 
the SUB$n\hspace{-0.1cm}-\hspace{-0.1cm}n$ approximation with $n=\{2,4,6\}$. 
Note that $N_F$ indicates the number of fundamental clusters
at each level of approximation. For comparison we present
 the results of the second-order spin wave theory (SWT) \cite{Hamer}. } 
\begin{center}
\begin{tabular}{|c|c|c|c|c|c|c|c|c|} \hline 
         &\multicolumn{3}{|c|}{$S = 3/2$}
        &\multicolumn{3}{|c|}{$S = 2$}\\ \hline  
&$N_F$  &$E_g/N$       &$M/S$ 
&$N_F$  &$E_g/N$        &$M/S$ \\ \hline
SUB2-2      &1     &$-$4.943927   &0.936174  
                   &1     &$-$8.593510    &0.950368 \\ \hline
SUB4-4      &15     &$-$4.976427   &0.910266
                  &15     &$-$8.633108    & 0.93109\\ \hline
SUB6-6      &461  &$-$4.982685    &0.89816 \ 
                    &461  &$-$8.640356    &0.922284  \\ \hline
SUB$\infty$ &--      &$-$4.9878   &0.8687 
                        &--      &$-$8.6461       &0.9011     \\ \hline
SWT          
                &--    &$-$4.9862      &0.8692     
                 &--    &$-$8.6442       &0.9018       \\ \hline
\end{tabular}
\end{center}
\label{tab1}
\end{table}

Finally, we consider spin $s=3/2$ and $s=2$. The results for the extrapolated
sublattice magnetization for spin values $s=1/2, 1, 3/2, 2$ using 
SUB$n\hspace{-0.1cm}-\hspace{-0.1cm}n$ approximation  for $n=2,4,6$
are shown in Fig.~\ref{fig4}. Evidently the sublattice magnetization $M/s$
increases with $s$ demonstrating the decreasing influence of
quantum fluctuations with growing spin quantum number. 
The critical value for $s=3/2$ 
is obtained as $J'_c \approx 18.5$ which is again too large
in comparison to the MFA result. The extrapolation of the inflection points 
leads to $J'_c \approx 10.9$. 
Note that we have calculated $M$ using the CCM up to $J'=100$ 
for $s=2$. However, we were unable to find a vanishing $M$ 
(i.e.,  the critical value $J'_c $ obtained by extrapolation of 
the order parameter would be larger than $100$.) 
Results for the point of inflection of $M$ were similarly 
contradictory, and so the results for the position of the 
phase transition point predicted by the CCM for $s=2$ are 
not included here.

We conclude that the CCM SUB$n\hspace{-0.1cm}
-\hspace{-0.1cm}n$ approximation is inappropriate in 
order to describe the quantum phase transition correctly  for 
higher spin values (namely, $s > 3/2$) at the levels of 
approximation currently available for present-day computers.  
However, we do observe that the tendency for critical 
value $J'_c$ to increase with growing $s$ is observed using 
the CCM for $s \le 3/2$, as expected.
This problem of reliability might be resolved
by going to higher orders of truncation index $m$,
although we note that the computational problem
is very difficult (e.g., with $N_F = 108033$ for SUB8-8 for
$s=3/2$) and so this is not considered here. 
We note that LSUB$n$ calculations do not place 
a restriction on the total number of spin
flips used in the CCM correlation 
operators, although the fundamental clusters 
are restricted to remain within a locale 
defined by $n$. However, this again leads
to an extremely large number of fundamental clusters
even for low values of $n$ and for higher spin
quantum numbers, and so LSUB$n$ is not considered
here. Mean-field model states (e.g., based on 
the variational states in Sec. 3) might also
provide enhanced results for the CCM. 

As a byproduct we also present in Table \ref{tab1} the 
results for the sublattice magnetization $M/s$ 
for higher spin values for the pure square lattices ($J'=1$),
which are so far not calculated within CCM. We point out that for the
pure square lattice the results for $M$ are expected to be quite reliable,
since the true quantum ground state is close to the reference state used as
starting point. This is indeed confirmed by comparison with high precision
second-order spin wave results \cite{Hamer} also presented in Table \ref{tab1}.        
We mention that due the reduced symmetry the number of 
fundamental configurations increases in case of $J' \ne J$. For 
SUB$6\hspace{-0.1cm}-\hspace{-0.1cm}6$ we find $N_F=267, 1420, 1744, 1744$ 
for $s=1/2,1,3/2,2$, respectively.  Note that $N_F$ for $s=3/2$ and $s=2$
is equal only up to SUB$6\hspace{-0.1cm}-\hspace{-0.1cm}6$ but differs 
for higher levels of approximation.  
For completeness we also give  the sublattice magnetization for $s=1/2$: 
$M/s=0.63$ (note that this value can
be improved by considering also SUB$8\hspace{-0.1cm}-\hspace{-0.1cm}8$
for the extrapolation, which yields $M/s= 0.62$ \cite{bishop00}) and for
$s=1$: $M=0.81$ (see also
\cite{farnell01}). 
\begin{figure}[ht]
\begin{center}\epsfxsize=30pc
\epsfbox{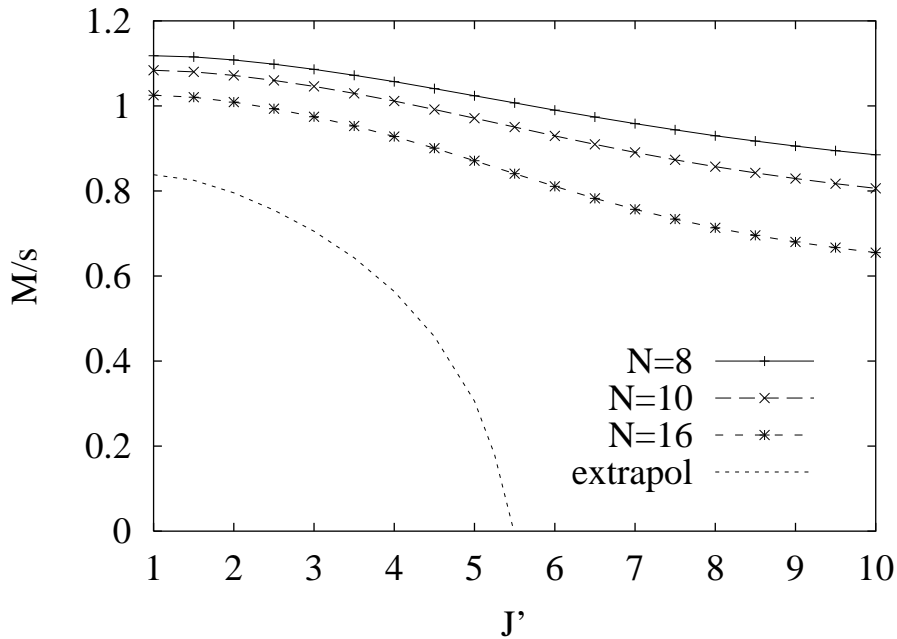}
\end{center}
\caption{\label{fig5} Sublattcie magnetization $M/s$ versus $J'$
for spin quantum number $s=1$ using exact diagonalization of finite lattices of $N=8, 10, 16$, 
see text.}
\end{figure}  

\section{Exact diagonalization (ED)}\label{ed}
In addition to the MFA and the CCM we use 
the exact diagonalization Lanczos technique 
to calculate the order parameter for finite square lattices with 
periodical boundary conditions. 
The calculations are performed using the
J.Schulenburg's {\it spinpack} \cite{spinpack}.
As usual for ED (see, e.g. Ref. \cite{wir04})
we calculate the square of the sublattice magnetization $M^2$ defined by 
$M^2 =
\langle [ \frac{1}{N}\sum^N_{i=1}\tau_i{\bf s}_i ]^2 \rangle 
$ with the staggered factor $\tau_{i \in A} =+1$, $\tau_{i \in B} =-1 $. 
For the finite-size 
scaling of $M^2$
we use the standard three-parameter 
formula \cite{neuberger89,hasenfratz93,wir04}
$M^2(N) =  M^2(\infty)+ c_1N^{-1/2} + c_2N^{-1}$.
The critical value $J'_c$ is that point where
$M^2(\infty)$ vanishes.
Again we are faced with the problem, that the method becomes less reliable 
for larger quantum numbers $s$.
While for $s=1/2$ one can calculate the GS for the $J$-$J'$ model up to
$N=32$ \cite{krueger00,tomczak01,rachid04}
sites one is restricted  to lattices of up to $N=20$ for $s=1$, 
up to $N=16$ for $s=3/2$ and up to $N=10$
for $s=3/2$. 
Since for $s=2$ we have only  two lattices ($N=10$ and $N=8$) 
with the full lattice symmetry,  we do not consider $s=2$ within ED. 
To treat all three cases  in a consistent way 
we consider only $N = 8, 10, 16$ for $s=1/2, 1, 3/2$. It is clear that the
resulting finite-size extrapolation remains quite poor and allows only some
qualitative conclusions.     
We present for illustration the results for the
order parameter $M$ for $s=1$ in Fig. \ref{fig5}.
The critical values obtained by finite size extrapolation of $M$ are:
$J'_c \approx 2.2$ for $s=1/2$, $J'_c \approx 5.5$ for $s=1$ and 
$J'_c \approx 10.1$ for $s=3/2$.
These data confirm the tendency found by MFA and CCM that the increase of 
$J'_c$ with $s$ is stronger than linear.   
\begin{figure}[ht]
\begin{center}\epsfxsize=30pc
\epsfbox{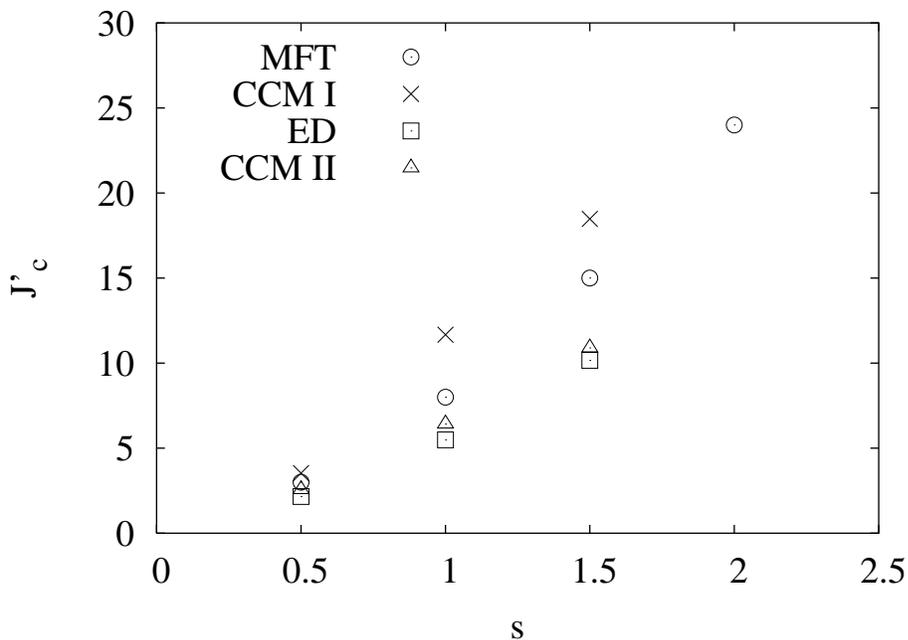}
\end{center}\caption{\label{fig6} The 
critical value $J'_c$  versus spin quantum number 
$s$ obtained by different methods. MFA: 
variational mean-field approach (see Sec. \ref{mfa}); CCM I: 
coupled cluster method (extrapolation of the order parameter, see Sec. 
\ref{ccm});
CCM II: 
coupled cluster method (extrapolation of the inflection point, see Sec. 
\ref{ccm});
ED: 
exact diagonalization (see Sec. 
\ref{ed}).
}
\end{figure}

\section{Summary and Discussion}

We have investigated the ground-state magnetic order  
parameter for the square-lattice isotropic Heisenberg  
antiferromagnet with two kinds of nearest-neighbor 
exchange bonds ($J$-$J'$ model) by using a 
variational mean-field approach (MFA), the coupled cluster
method (CCM) and exact diagonalizations (ED).  
In particular, we have studied the influence of 
the spin quantum number $s$ on the quantum
critical point $J'_c$. Our results for $J'_c$ 
are presented  in Fig.\ref{fig6}, and we note that 
a transition from a semi-classically N\'eel ordered phase 
to a magnetically disordered phase occurs at
this point. Obviously, there is an increase of 
$J'_c$ with $s$ signaling the diminishing of 
quantum effects. We have presented evidence that 
the critical  value $J'_c$ increases with
growing $s$ according to $J'_c \propto s(s+1)$.

\ack
We  thank the DFG for financial support (project Ri 614/14-1).
The authors are indebted to J. Schulenburg for assistance in numerical
calculations.

\section*{References}

\end{document}